\documentclass[11pt]{article} 
\pdfoutput=1
\usepackage{amssymb,graphicx} 
\usepackage{epstopdf}
\usepackage{amsmath,amsfonts}
\usepackage{epsfig} 
\usepackage{graphicx,graphics}
\usepackage[dvipsnames]{xcolor}
\usepackage[utf8]{inputenc}
\usepackage{hyperref}

\begin{document} 

\sloppy

\title{\bf Shrouded black holes in Einstein-Gauss-Bonnet gravity}
\author{Eugeny Babichev$^{a}$, William T.~Emond$^{b}$, Sabir Ramazanov$^b$\\
\small{\em $^b$Universit\'e Paris-Saclay, CNRS/IN2P3, IJCLab, 91405 Orsay, France}\\
\small{\em $^b$CEICO, FZU-Institute of Physics of the Czech Academy of Sciences,}\\
\small{\em Na Slovance 1999/2, 182 21 Prague 8, Czech Republic}
}

{\let\newpage\relax\maketitle}

{\let\newpage\relax\maketitle}

\begin{abstract} 

We study black holes in a modified gravity scenario involving a scalar field quadratically coupled to the Gauss-Bonnet invariant. The scalar is assumed to be in a spontaneously broken phase at spatial infinity due to a bare Higgs-like potential. For a proper choice of  sign, the non-minimal coupling to gravity leads to symmetry restoration near the black hole horizon, prompting the development of the scalar wall in its vicinity. 
The wall thickness depends on the bare mass of the scalar and can be much smaller than the Schwarzschild radius. 
In a weakly coupled regime, the quadratic coupling to the Gauss-Bonnet invariant effectively becomes linear,  and no walls are formed.
We find approximate analytical solutions for the scalar field in the test field regime, and obtain numerically static black hole solutions within this setup.  
We discuss cosmological implications of the model and show that it is fully consistent with the existence of an inflationary 
stage, unlike the spontaneous scalarization scenario assuming the opposite sign of the non-minimal coupling to gravity. Our model predicts the speed of gravitational waves to be extremely close 
to unity, -- in a comfortable agreement with the observation of the GW170817 event 
and its electromagnetic counterpart.

\end{abstract}

\sloppy

\section{Introduction}

The growing number of black holes discovered by LIGO, VIRGO, and KAGRA collaborations~\cite{LIGOScientific:2016aoc,LIGOScientific:2018mvr,LIGOScientific:2020ibl,LIGOScientific:2021djp} greatly improves the possibilities for strong field tests of gravity, which have been mainly limited to binary pulsars until the recent past. Such strong field tests are crucial, if a given modification of gravity is indistinguishable from General Relativity (GR) in the weak field regime, i.e., in the Solar system.

In this work we focus on scalar-tensor theories of gravity involving 
a scalar field $\varphi$ coupled to the Gauss--Bonnet invariant ${\cal R}^2_{\mathrm{GB}}$. Such couplings have received a lot of attention in the literature: in particular, black hole solutions have been identified for a dilatonic coupling of the form $\sim e^{\varphi} {\cal R}^2_{\mathrm{GB}}$~\cite{Mignemi:1992nt, Kanti:1995vq, Torii:1996yi} as well as for the linear coupling $\sim \varphi {\cal R}^2_{\mathrm{GB}}$~\cite{Sotiriou:2013qea, Sotiriou:2014pfa,Babichev:2016rlq,Creminelli:2020lxn}. Recently, it has been shown that the quadratic scalar-Gauss-Bonnet coupling $\sim \varphi^2 {\cal R}^2_{\mathrm{GB}}$ leads to spontaneous scalarization of black holes and neutron stars for a certain choice of the sign of the coupling~\cite{Doneva:2017bvd, Antoniou:2017acq, Antoniou:2017hxj, Silva:2017uqg}. In a nutshell, scalarization occurs due to the instability of the trivial solution $\varphi=0$ reproducing GR. As a result, the scalar acquires a non-trivial profile in the vicinity of compact objects, which in turn provokes potentially testable departures from GR. Note that historically the first example of spontaneous scalarization has been given in Ref.~\cite{Damour:1993hw}, assuming the scalar coupling to the Ricci curvature, in which case scalarization takes place only for neutron stars. This model of scalarization is subject to stringent observational constraints~\cite{Freire:2012mg} clearly demonstrating how 
deviations from GR caused by scalarization can be efficiently tested. Implications of scalar-Gauss-Bonnet coupling for black hole mergers~\cite{Witek:2018dmd, Silva:2020omi, East:2021bqk, Elley:2022ept} and binary pulsars~\cite{Danchev:2021tew} are currently under active investigation.

We find new solutions for static black holes in a model with the quadratic coupling $\sim \varphi^2 {\cal R}^2_{\mathrm{GB}}$ assuming a non-trivial bare (i.e., non-vanishing upon switching off gravity) potential $V(\varphi)$. Studies of scalarization 
performed so far do not include the potential $V(\varphi)$, or assume that it has one trivial minimum, which can be set at the origin $\varphi=0$ with no loss of generality~\cite{Macedo:2019sem, Doneva:2019vuh, Hod:2019vut, Ripley:2020vpk, Staykov:2021dcj}. Unlike the previous studies, in this paper we consider a bare potential $V(\varphi)$, which spontaneously breaks $Z_2$-symmetry, i.e., $V(\varphi) \propto (\varphi^2 -v^2)^2$. Here 
$v$ is the expectation value of the scalar field $\varphi$ at spatial infinity\footnote{Note that Refs.~\cite{Antoniou:2017acq, Antoniou:2017hxj, Papageorgiou:2022umj} assume ad hoc a non-trivial background value of the field $\varphi$ without introducing the potential $V(\varphi)$.}. For a proper choice of the sign, the role of the scalar-Gauss-Bonnet coupling is to restore the symmetry in the vicinity of the black hole. 
Consequently, for a sufficiently large coupling to the Gauss-Bonnet invariant, the scalar $\varphi$ sits close to zero near the horizon, and the overall scalar profile resembles that of a domain wall, which becomes 
more pronounced as one increases the bare mass of the field $\varphi$. We confirm this picture of shrouded black holes by running the full system of equations of motion involving the scalar field and metric potentials in Section~4. Before that, in Section~3 
we treat the scalar profile in the test field approximation, in which case not only do numerical simulations become less cumbersome, but one can also estimate the profile analytically. In a weak coupling regime, 
the departure of the scalar from its expectation value $v$ is always small, and one effectively recovers the model with a linear coupling to the Gauss-Bonnet invariant. 

In our case, the sign of the scalar-Gauss-Bonnet coupling is opposite compared to the scenario without the symmetry breaking potential~\cite{Doneva:2017bvd, Silva:2017uqg}\footnote{Here we would like to stress that we consider static black holes. In the case with $V(\varphi)=0$, scalarization of Kerr black holes may occur for either sign of the coupling depending on the spin~\cite{Dima:2020yac, Berti:2020kgk, Herdeiro:2020wei}, see also Ref.~\cite{Cunha:2019dwb}.}.. 
This is advantageous in view of the following problem: in the spontaneous scalarization scenario of Refs.~\cite{Doneva:2017bvd, Silva:2017uqg} the scalar-Gauss-Bonnet coupling mimics a large tachyonic mass in the accelerated Universe leading to a runaway solution inconsistent with existence of inflation~\cite{Anson:2019uto}. 
With our choice of the sign of the coupling no instability during inflation occurs. Nevertheless, we face possible issues with post-inflationary cosmological evolution, 
albeit significantly alleviated. Indeed as we show in Section~5 there is an instability during the radiation-dominated stage. This instability, however, develops very slowly and thus does not spoil standard cosmological picture, provided that the reheating temperature is bounded as $T_{\mathrm{reh}} \lesssim 1~\mbox{GeV}$. In Section~5, we also discuss 
stringent limits on model parameters following from the possibility of domain wall formation and Dark Matter overproduction, and point out the ways of avoiding these constraints. Finally, we show there that the speed of gravitational waves is well  within the existing bound imposed from the simultaneous observation of the gravitational wave signal GW170817~\cite{LIGOScientific:2017vwq} and its electromagnetic counterpart~\cite{Goldstein:2017mmi}.

\section{The model}

We consider the following scalar-tensor model: 
\begin{equation}
\label{action}
S=\int d^4 x \sqrt{-g} \left[\frac{1}{2} R-\frac{1}{2}\nabla_{\mu} \varphi \nabla^{\mu} \varphi-V(\varphi) - \ell^2 \varphi^2 {\cal R}^2_{\mathrm{GB}} \right] \; , 
\end{equation}
where ${\cal R}^2_{\mathrm{GB}}=R^2-4R_{\mu \nu} R^{\mu \nu} +R_{\mu \nu \lambda \rho} R^{\mu \nu \lambda \rho}$ is the Gauss--Bonnet invariant and $\ell$ is a parameter with dimension of length. 
We work in units $8\pi G_N=1$, where $G_N$ is the Newton's constant, and we assume the mostly positive metric signature $(-,+,+.+)$. The bare scalar potential $V(\varphi)$ is taken to be of the form
\begin{equation}
\label{scalarpotential}
V(\varphi)=\frac{h^2}{4} \cdot \left(\varphi^2-v^2 \right)^2, \qquad v=\frac{\mu}{h}\; ,
\end{equation}
where $v$ is a vacuum expectation value of the field $\varphi$, $h^2$ is the quartic self-interaction coupling constant, and $-\mu^2$ is the tachyonic bare mass squared of the field $\varphi$. 
We assume that the scalar $\varphi$ resides in the minimum of its potential $V(\varphi)$ spontaneously breaking $Z_2$-symmetry, 
i.e., 
\begin{equation}
\label{scalarinfinity}
\varphi_{\infty} =v \; .
\end{equation}
(We choose the minimum with a positive sign with no loss of generality.) Note that the potential~\eqref{scalarpotential} fulfils $V(\varphi_{\infty})=0$, so that it does not contribute to the cosmological constant $\Lambda$ (cf. Ref.~\cite{Bakopoulos:2018nui}). Neglecting $\Lambda$, 
we therefore have asymptotically Minkowski spacetime. In this work, we assume a particular sign of the scalar-Gauss-Bonnet coupling:
\begin{equation}
\label{coupling}
\ell^2>0 \; .
\end{equation}
With this choice, which is opposite compared to that of Refs.~\cite{Doneva:2017bvd, Silva:2017uqg}, $Z_2$-symmetry, spontaneously broken at spatial infinity, can be restored in the strong gravity regime, in the vicinity of a compact object. 
Once this regime is realized, the scalar field acquires a non-trivial configuration in the form of a shell that shrouds a black hole.

The main goal in this work is to find static black hole solutions in the model~\eqref{action}. Phenomenology of the model in the weak field regime, i.e., in the Solar system, has been considered in Ref.~\cite{Amendola:2007ni}. The bound on the model parameters inferred there using the Shapiro time delay measurements by the Cassini spacecraft reads $\ell^2 v \lesssim 1.6 \cdot 10^{20}~\mbox{m}^2$. This bound is very weak allowing $\ell$ to be orders of magnitude above the Schwarzschild radius of Solar mass objects.

We are interested in the situation where the scalar-Gauss-Bonnet coupling, negligible at spatial infinity, eventually overcomes the bare mass of the field $\varphi$, as one approaches the horizon. Otherwise, the field $\varphi$ remains in the spontaneously 
broken phase. It is convenient to define the crossover radius $r_{\mathrm{cross}}$, where the scalar-Gauss-Bonnet coupling is of the order of the bare mass term $\sim \mu^2 \varphi^2$, namely
\begin{equation}
\mu^2 \sim 2 \ell^2 {\cal R}^2_{\mathrm{GB}} (r_{\mathrm{cross}}) \; .
\end{equation}
Given that ${\cal R}^2_{\mathrm{GB}} (r) \approx 12 r^2_S/r^6$ for $r \gg r_S$, we get 
\begin{equation}
\label{turn}
r_{\mathrm{cross}} \sim 1.7\cdot \left(\frac{\ell r_S}{\mu} \right)^{1/3} \; .
\end{equation}
Hereafter, the subscript $'S'$ denotes values on the Schwarzschild radius. Formation of a scalar wall is possible provided that $r_{\mathrm{cross}} \gtrsim r_S$, which translates into the bound on $\mu$:
\begin{equation}
\label{upperbound}
\mu \lesssim \frac{5\ell}{r^2_S} \; ,
\end{equation}
or
\begin{equation}
\label{upperconcrete}
\mu \lesssim  \frac{3  \ell}{r_S} \cdot \frac{M_{\odot}}{M} \cdot 10^{-10}~\mbox{eV} \; .
\end{equation}
Note that for Solar mass objects and $\ell$ not much larger than $r_S$, the scalar $\varphi$ is ultra-light. So small masses $\mu$ are not at all exotic in physics, e.g., they are characteristic for axions and axion-like particles.

\section{Test field approximation}

Let us start by constructing a solution for the scalar field $\varphi$ in the test field approximation, i.e., neglecting deviations of gravitational potentials from their GR values. 
Such a simplification is justified in theoretically well-motivated ranges of parameters $\ell \ll r_S$ (perturbative regime) and/or $v \ll 1$. Furthermore, we will see that certain non-perturbative effects, which may have non-trivial phenomenological 
consequences, arise already in the test field approximation.  

As we focus on static black holes in the present work, the line element in Schwarzschild coordinates is given by 
\begin{equation}
\label{spherical}
ds^2 =-e^{\nu}dt^2 +e^{\lambda} dr^2 +r^2 \cdot \left(d\theta^2 +\sin^2 \theta d\varphi^2 \right) \; .
\end{equation}
GR values of the gravitational potentials $\nu$ and $\lambda$ are inferred from 
\begin{equation}
e^{\nu} =e^{-\lambda}=1-\frac{r_S}{r} \; .
\end{equation}
In the test field approximation, the equation of motion for the scalar $\varphi$ reads
\begin{equation}
\label{testequation}
\varphi'' +\left( \nu'+\frac{2}{r} \right) \cdot  \varphi' +\frac{8\ell^2 \varphi}{r^2} \cdot \left[\left(1-e^{\nu} \right) \cdot \left(\nu''+\nu'^{2}  \right) -e^{\nu} \cdot \nu'^{2} \right] -e^{-\nu} \cdot V_{,\varphi} =0 \; ,
\end{equation}
where $V_{,\varphi} \equiv \frac{d V}{d\varphi}$ and the prime denotes the derivative with respect to the radial coordinate $r$.
Note that Eq.~(\ref{testequation}) is invariant under the change
\begin{equation}
\varphi \rightarrow L \varphi, \qquad v \rightarrow L v, \qquad h \rightarrow \frac{h}{L} \qquad \left( \mu \rightarrow \mu \right), \qquad \ell \to \ell,
\end{equation}
where $L$ is a constant.
The regularity of the solution on the boundary requires that all the terms $\sim 1/ \left(1-\frac{r_S}{r} \right)$ in Eq.~\eqref{testequation} cancel out. 
This gives 
\begin{equation}
\label{testboundary}
 \varphi'_S -\frac{24\ell^2}{r^3_S} \cdot \varphi_S -r_S V_{,\varphi} (\varphi_S) =0\; .
\end{equation}
At infinity the boundary condition for the field $\varphi$ is given by Eq.~\eqref{scalarinfinity}. 

\begin{figure}[tb!]
  \begin{center}
    \includegraphics[width=\columnwidth,angle=0]{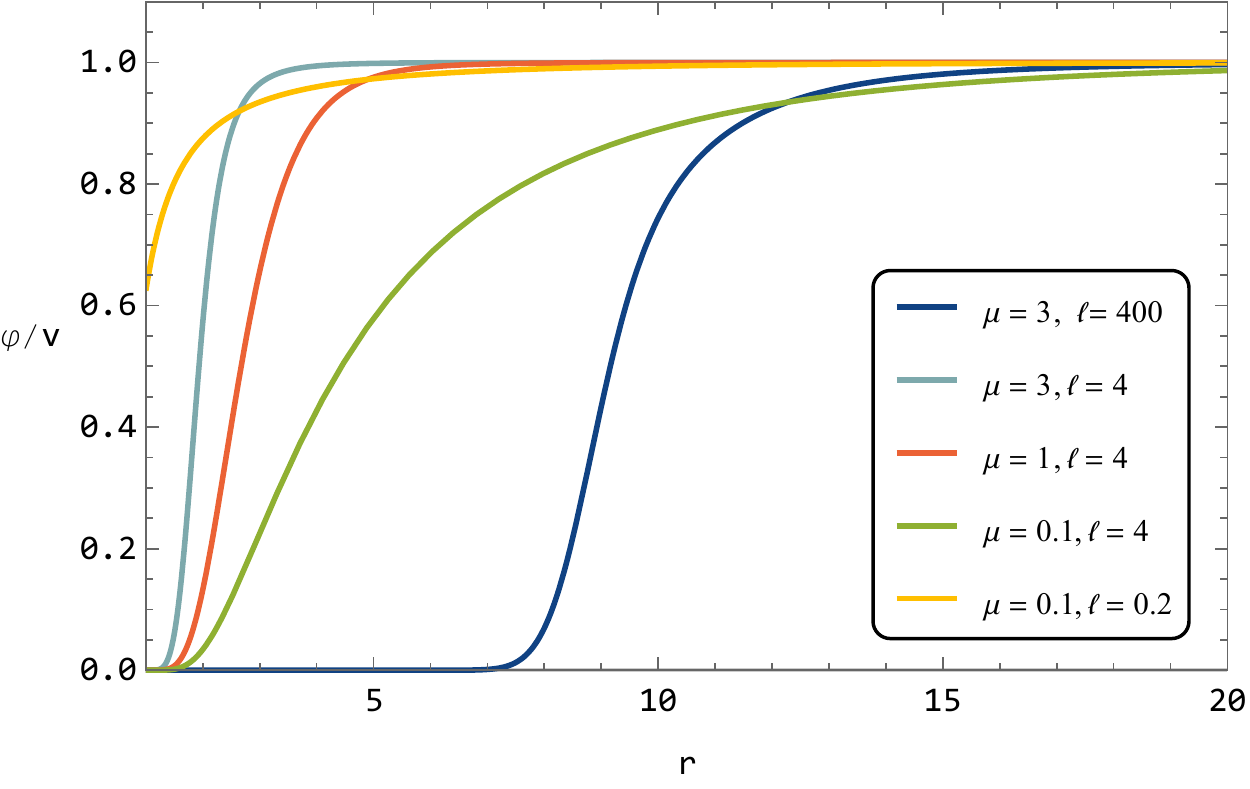}
  \caption{Scalar field $\varphi$ profiles around static black holes obtained in the test field approximation are shown for different sets of model parameters in dimensionless variables~\eqref{dimensionless}. The scalar field is normalized to its expectation value $v$ at spatial infinity. The sets of parameters are chosen to cover the three cases outlined in Eq.~\eqref{cases}: the orange line corresponds to case I (perturbative regime), the green line stands for case II, where the walls first appear; the walls become steeper in the large $\mu$ regime corresponding to case III depicted with the dark blue, aquamarine, and red lines.
  }\label{Test_field}
  \end{center}
\end{figure}

\begin{figure}[tb!]
  \begin{center}
        \includegraphics[width=\columnwidth, angle=0]{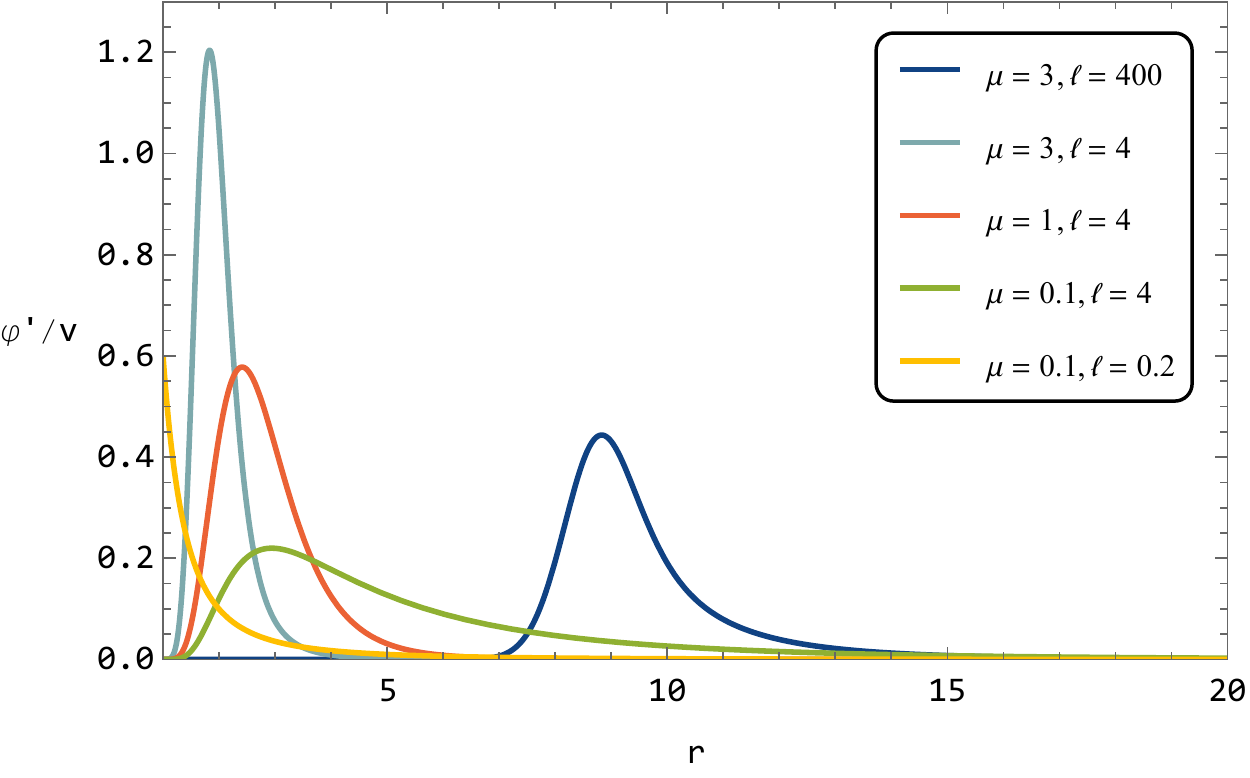}
  \caption{The first derivative with respect to normalized scalar profiles of Fig.~\ref{Test_field} is shown in dimensionless variables~\eqref{dimensionless}. In the large $\ell$ regime, the scalar field derivative has peaks, which match the locations of the walls shown in~Fig.~\ref{Test_field}; a wall width is estimated as a peak width. For increasing $\mu$ and fixed $\ell$, a wall moves towards a horizon and its width shrinks, in accordance with Eqs.~\eqref{location} and~\eqref{width}. }\label{Derivative_test}
  \end{center}
\end{figure}

We solve Eq.~\eqref{testequation} with the boundary conditions~\eqref{scalarinfinity} and~\eqref{testboundary} using both the shooting method and the relaxation technique. The results are shown in Figs.~\ref{Test_field} and~\ref{Derivative_test} in terms of the field $\varphi$ and its derivative $\varphi'$, respectively. At this point, we have switched to dimensionless variables:
\begin{equation}
\label{dimensionless}
r \rightarrow \frac{r}{r_S} \qquad \varphi \rightarrow \varphi \qquad \ell \rightarrow \frac{\ell}{r_S} \qquad \mu \rightarrow \mu \cdot r_S \qquad h \rightarrow h \cdot r_S  \; .
\end{equation}
Recall that we assume $8\pi G_N=1$. 
Qualitatively, the scalar field profile is different in three cases:
\begin{align}
\label{cases}  
\begin{cases}
\ell \ll r_S \qquad  \qquad \qquad \qquad \mbox{(Case~I)} \\
\ell \gg r_S \qquad \mu \ll \frac{1}{\sqrt{\ell r_S}} \qquad \mbox{(Case~II)}\\
\ell \gg r_S \qquad \mu \gg \frac{1}{\sqrt{\ell r_S}} \qquad \mbox{(Case~III)}
\end{cases}
\,.
\end{align}
In all three cases, we assume that the crossover radius is larger than the Schwarzschild radius, so that Eq.~\eqref{upperbound} is fulfilled. 

Case I corresponds to the perturbative regime (with $\ell^2$ being a perturbative parameter), i.e., the field $\varphi$ remains close to its bare expectation value $v$ all the way from 
infinity till the horizon surface, and one has $\varphi \rightarrow v$ in the limit $\ell \rightarrow 0$. Thus, for $\ell \ll r_S$, one can effectively replace the quadratic coupling $\sim \varphi^2 {\cal R}^2_{\mathrm{GB}}$ with a linear one $\sim v \varphi {\cal R}^2_{\mathrm{GB}}$. In cases II and III, there is a non-perturbative dependence of the scalar field solutions on the coupling constant $\ell^2$, manifested in particular by the strong exponential suppression of the field $\varphi$ near the horizon surface. This confirms the main idea underlying the model: for sufficiently large $\ell^2$, the scalar-Gauss-Bonnet coupling leads to the symmetry restoration in the vicinity of a black hole. As a result, we observe a wall -- the scalar field configuration interpolating between the broken and unbroken phase at spatial infinity and near the horizon, respectively. 
Cases II and III differ by the ratio of the crossover radius $r_{\mathrm{cross}}$ and $r \sim \mu^{-1}$. Namely, the sphere $r \sim r_{\mathrm{cross}}$ is inside the sphere $r \sim \mu^{-1}$ in case II, and vice versa in case III. In practice, this difference means that the walls are much steeper in case III. As $\mu$ increases, the wall becomes sharper and moves closer to the horizon.

In the remainder of this section, we discuss the analytical solution for the field $\varphi$ in the test field regime. Details of the derivation of this solution can be found in the Appendix, and below we summarize the main results.

{\it Case I}. For $\ell \ll r_S$, the departure of the field $\varphi$ (including its value at the horizon $\varphi_S$) from the expectation value at inifnity $v$ is very small and hence the effect of the Gauss-Bonnet term is rather mild, -- in an agreement with the numerical result shown in Figs.~\ref{Test_field} and~\ref{Derivative_test}. In particular, the scalar field profile at distances $r \gg r_S$ is described by
\begin{equation}
\label{profilecaseI}
\varphi \simeq v \cdot \left(1-\frac{10 A \ell^2}{r_S r} e^{-\sqrt{2} \mu r} \right) \; ,
\end{equation}
where $A$ is an order one constant. Note that the coupling constant $\ell^2$ enters the expression~\eqref{profilecaseI} in a perturbative fashion. This confirms  that the regime $\ell \ll r_S$ corresponds to the weak field limit. 

{\it Case II.} In the limit of large $\ell$ and relatively small $\mu$, the scalar has an exponentially suppressed value at the horizon $\varphi_S$. At relatively small values of $r$, the scalar exponentially grows as
\begin{equation}
\label{profileinterII}
\varphi \simeq  0.3 \cdot v \cdot \left(\frac{r^2}{ \sqrt{6} \ell r_S} \right)^{1/4} \cdot e^{-\frac{\sqrt{6} \ell r_S}{r^2}}  \qquad \left(r_S \ll r \ll \sqrt{\ell r_S} \right) \; .
\end{equation}
We observe a non-perturbative dependence of the scalar profile on the coupling $\ell^2$.
At $r \sim \sqrt{\ell r_S}$ the exponential growth transitions to an exponentially slow relaxation towards the vacuum value $v$ at infinity,
\begin{equation}
\label{profilelargeII}
\varphi \simeq v \cdot \left[1-  \frac{3\sqrt{\ell r_S}}{2 r}  \cdot e^{-\sqrt{2} \mu r} \right]  \qquad \left(r \gg \sqrt{\ell r_S} \right)\; .
\end{equation}
Overall, the scalar profile has a configuration of a wall, which for $r_S \ll r \ll \mu^{-1}$ can be described by the following analytical expression:
\begin{equation}
\label{walllargel}
\varphi \simeq \frac{v}{2} \cdot \left(\frac{\sqrt{6} \ell r_S}{r^2} \right)^{1/4} \cdot K_{1/4} \left(\frac{\sqrt{6} \ell r_S}{r^2} \right) \; ,
\end{equation}
where $K_{1/4}$ is a modified Bessel function of the order $1/4$ of the second kind.
Note that Eq.~(\ref{walllargel}) reduces to expressions~\eqref{profileinterII} and~\eqref{profilelargeII} in given regimes. The wall configuration can be characterized by its location $r_{\mathrm{wall}}$ and the width $\delta_{\mathrm{wall}}$. It is not difficult to see that both are estimated as $r_{\mathrm{wall}} \sim \delta_{\mathrm{wall}} \sim \sqrt{\ell r_S}$, since $\sqrt{\ell r_S}$ is the only dimensionful parameter entering the expression~\eqref{walllargel}.

{\it Case III.} A similar story occurs for relatively large $\mu$: the field $\varphi$ starts from a tiny value $\varphi_S$ at the horizon and grows exponentially for $r_{S} \lesssim r \lesssim r_{\mathrm{cross}}$,
\begin{equation}
\label{profileinterIII}
\varphi \simeq \frac{v}{2} \cdot \sqrt{\frac{r}{r_{\mathrm{cross}}}}  \cdot e^{-\frac{\sqrt{6} \ell r_S}{r^2} \cdot \left(1-\frac{r^2}{r^2_{\mathrm{cross}}} \right)}  \qquad \left(r_{S} \ll r \ll r_{\mathrm{cross}} \right) \; .
\end{equation}
For $r\gtrsim r_{\mathrm{cross}}$, the growth is shut off:
\begin{equation}
\label{profilelargeIII}
\varphi \simeq v \cdot \left[1- \frac{r_{\mathrm{cross}}}{2r} \cdot e^{\sqrt{2} \mu (r_{\mathrm{cross}}-r)}  \right] \qquad \left(r \gg r_{\mathrm{cross}} \right) \; .
\end{equation}
As $\mu$ increases, with $\ell$ held constant, the radius $r_{\mathrm{cross}}$ shrinks, and consequently the scalar profile becomes more and more shallow. 
Eventually, when $\mu$ hits the upper bound in Eq.~\eqref{upperbound}, one gets $\varphi \simeq v$. This could be expected from the beginning: 
when the bare potential $V(\varphi)$ dominates over the scalar-Gauss-Bonnet coupling term all the way down to the horizon surface, the field $\varphi$ remains in the broken phase.

 The configuration described by Eqs.~\eqref{profileinterIII} and~\eqref{profilelargeIII} corresponds to a wall, which is much steeper than in case II, -- in an agreement with the numerical results shown in Figs.~\ref{Test_field} and~\ref{Derivative_test}. The location of the wall coincides with the peak of the derivative $\varphi'$ of the scalar profile given by Eqs.~\eqref{profileinterIII} and~\eqref{profilelargeIII}: 
\begin{equation}
\label{location}
r_{\mathrm{wall}} \sim r_{\mathrm{cross}} \sim 1.7 \cdot \left(\frac{\ell r_S}{\mu} \right)^{1/3} \; ,
\end{equation}
see Eq.~\eqref{turn}. The naive estimate for the wall width $\delta_{\mathrm{wall}}$ obtained by extrapolating expressions~\eqref{profileinterIII} and~\eqref{profilelargeIII} to the wall center at $r \sim r_{\mathrm{cross}}$, reads $\delta_{\mathrm{wall}} \sim \mu^{-1}$. 
A more accurate analysis of the field behaviour around the location of the wall presented in the end of the Appendix gives a slightly different estimate: 
\begin{equation}
\label{width}
\delta_{\mathrm{wall}} \sim \frac{(\ell r_S)^{1/9}}{\mu^{7/9}} \; .
\end{equation}
Note that $\mu$ can reach values as large as $\mu \sim \ell/r^2_S$ (for larger values the wall is not formed). In this limiting case, the 
wall is located very close to the horizon surface, and its width is a small fraction of the Schwarzschild radius.

\section{Numerical analysis of the full system of equations}

\begin{figure}[tb!]
  \begin{center}
    \includegraphics[width=\columnwidth,angle=0]{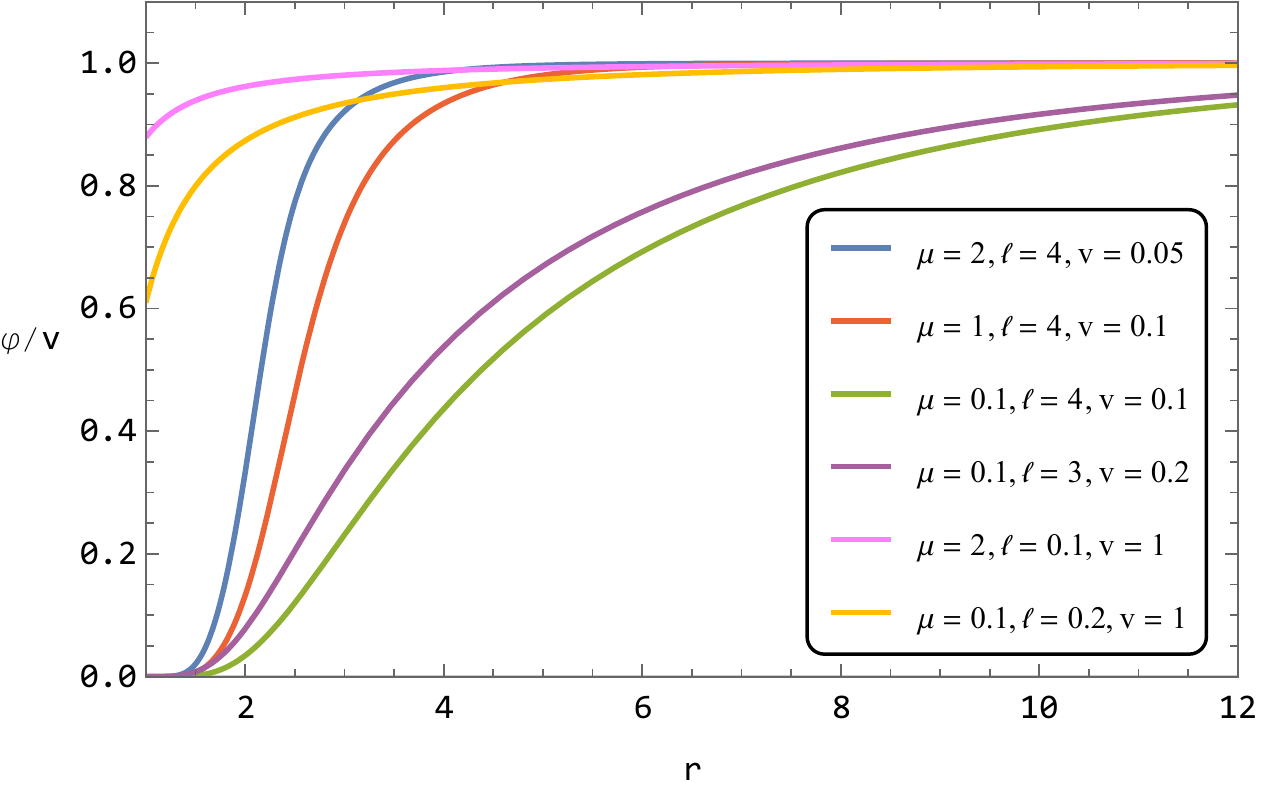}
  \caption{Scalar field profiles around static black holes in the model given by Eqs.~\eqref{action} and~\eqref{scalarpotential} obtained by numerically solving the full system of Eqs.~\eqref{tt},~\eqref{rr}, and~\eqref{eom} are shown for different sets of model parameters. The scalar field $\varphi$ is normalized to its expectation value $v$ at spatial infinity. Dimensionless variables of Eq.~\eqref{dimensionless} have been used.}\label{scalarprofile}
  \end{center}
\end{figure}

\begin{figure}[tb!]
  \begin{center}
    \includegraphics[width=\columnwidth,angle=0]{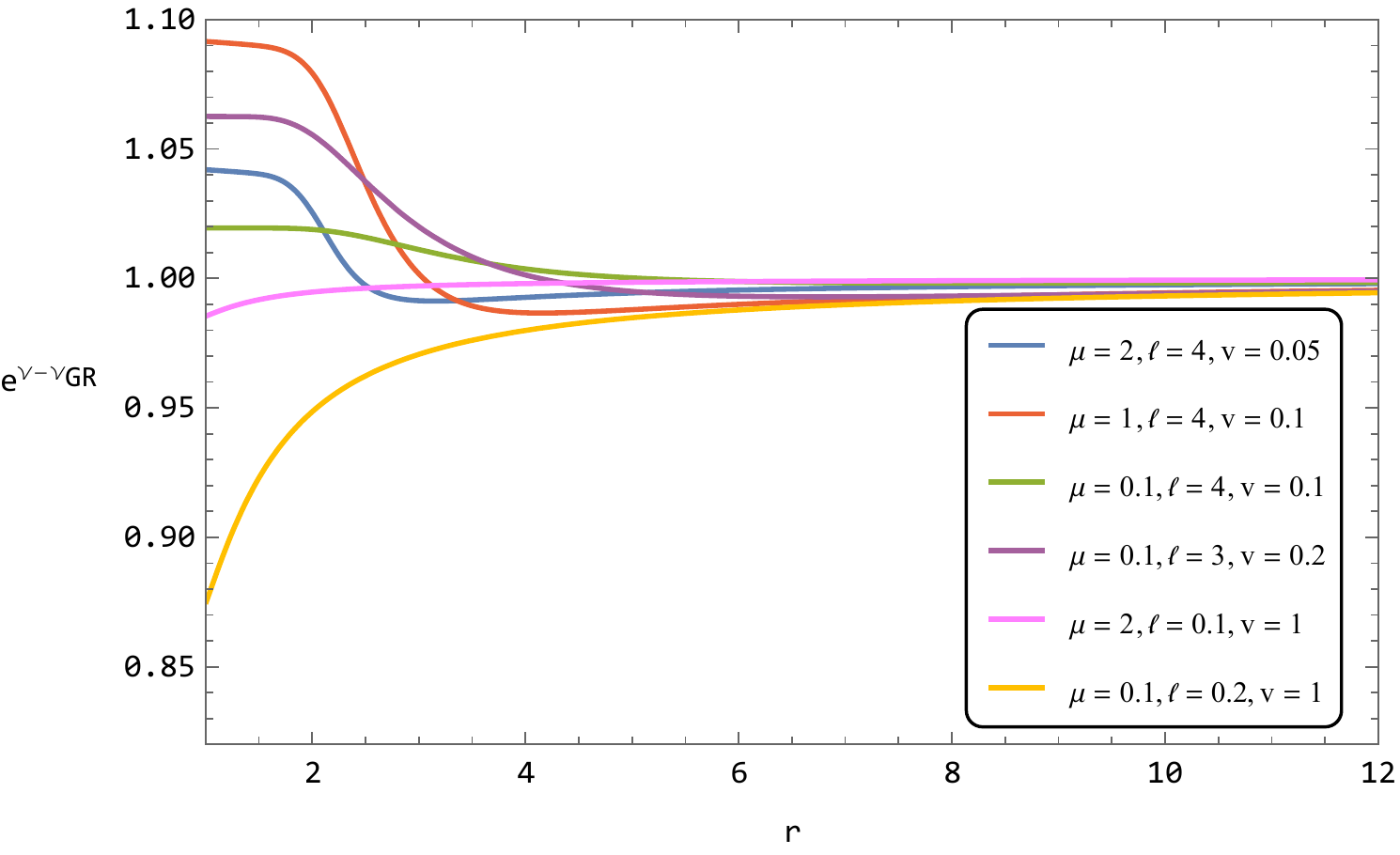}
  \caption{The metric function $e^{\nu}$ of static black holes in the model given by Eqs.~\eqref{action} and~\eqref{scalarpotential} is shown relative to its GR values $e^{\nu_{GR}}$ for the same sets of model parameters as in Fig.~\ref{scalarprofile}.}\label{metric}
  \end{center}
\end{figure}

\begin{figure}[tb!]
  \begin{center}
    \includegraphics[width=\columnwidth,angle=0]{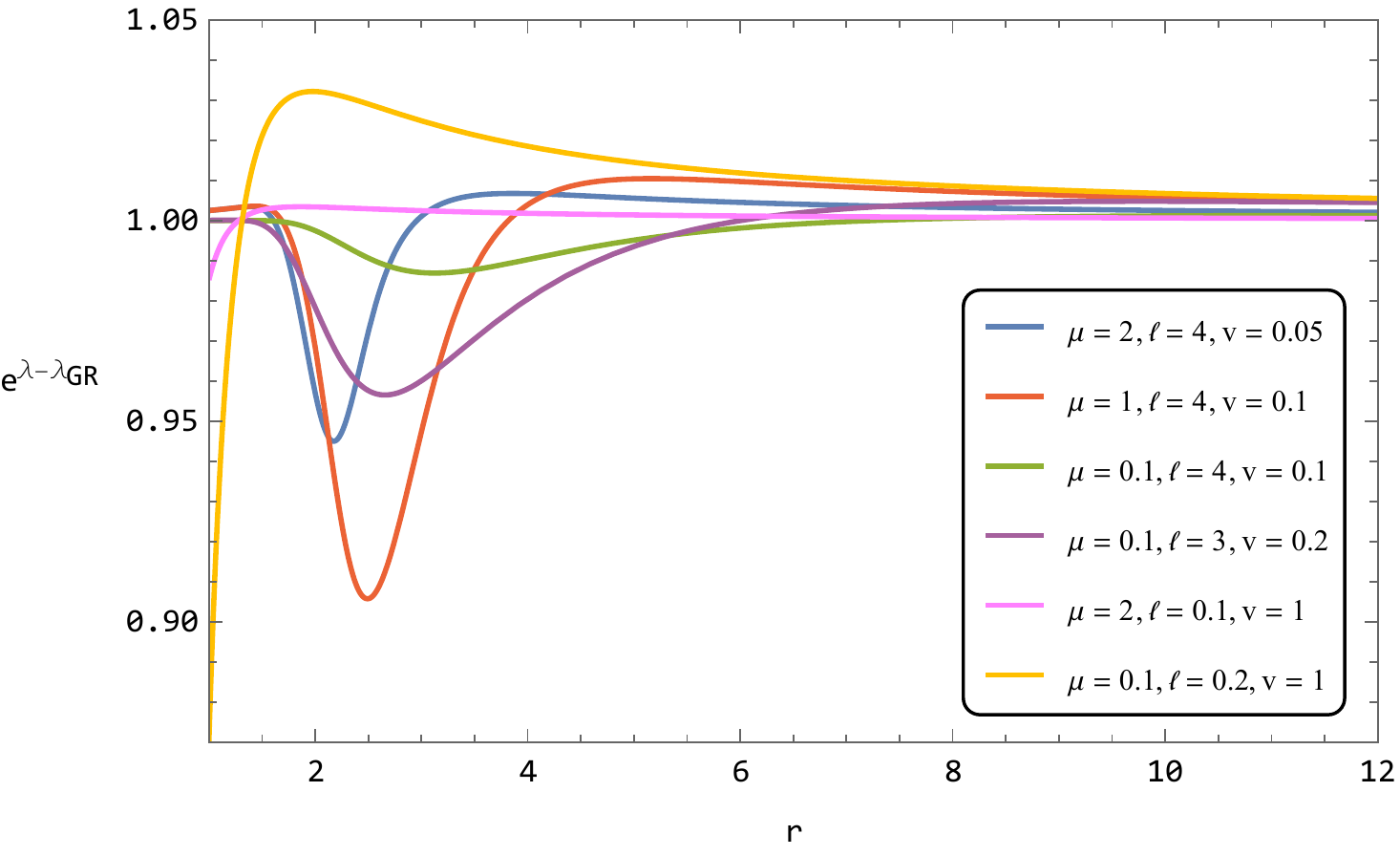}
  \caption{The metric function $e^{\lambda}$ relative to its GR values $e^{\lambda_{GR}}$ is shown for the same sets of parameters as in~Fig.~\ref{metric}.}\label{metric2}
  \end{center}
\end{figure}

In this section, we go beyond the test field approximation and numerically solve the full set of metric and the field $\varphi$ equations. As we are interested in static black hole solutions, we assume an ansatz for the metric of the form~\eqref{spherical}; however, in contrast to the previous section, here the metric is considered to be dynamical. For the metric potentials $\nu$ and $\lambda$ the equations of motion are given by
\begin{equation}
\label{tt}
(tt):~~~ \frac{1}{r} \left[1-\frac{8\ell^2}{r} (1-3e^{-\lambda}) \varphi \varphi' \right]  \lambda' +\frac{e^{\lambda} -1}{r^2} +\frac{16l^2 \cdot (1-e^{-\lambda})}{r^2} \varphi \varphi' -\frac{1}{2}\varphi'^2- e^{\lambda} V(\varphi)=0 \; ,
\end{equation}

\begin{equation}
\label{rr}
(rr):~~~ \frac{1}{r} \cdot \left[1-\frac{8\ell^2}{r} (1-3e^{-\lambda}) \varphi \varphi' \right] \nu' -\frac{e^{\lambda}-1}{r^2} -\frac{1}{2}\varphi'^2+ e^{\lambda} V(\varphi)=0 \; ,
\end{equation}
and 
\begin{equation}
\label{thetatheta}
\begin{split}
(\theta \theta )~~~ &\nu'' +\left(\frac{\nu'}2+\frac{1}{r} \right) \cdot \left(\nu'-\lambda' \right)-\frac{8\ell^2 \cdot e^{-\lambda}}{r} \cdot \left[3 \nu' \lambda'-2\nu''-\nu'^2 \right] \varphi \varphi'+\\&+\frac{16\ell^2 \cdot e^{-\lambda}}{r} \nu' \cdot \left(\varphi'^{2}+\varphi \varphi'' \right) +\varphi'^2 +2e^{\lambda} V(\varphi) =0 \; .
\end{split}
\end{equation}
The equation of motion for the field $\varphi$ reads
\begin{equation}
\label{eom}
\begin{split}
&\varphi'' +\left(\frac{\nu'-\lambda'}2 +\frac{2}{r} \right) \varphi' +\\&+
\frac{8\ell^2 \varphi }{r^2} \left[ (1-e^{-\lambda}) \cdot \left(\nu''+ \frac{\nu'-\lambda'}2\nu' \right)+e^{-\lambda} \nu' \lambda'  \right] -e^{\lambda} V_{,\varphi}=0 \; .
\end{split}
\end{equation}
Note that only three of the above four equations are independent due to the Bianchi identities.

Boundary conditions for gravitational potentials $\nu$ and $\lambda$ at the horizon surface are given by
\begin{equation}
\label{boundarypotentials}
   e^{\nu} \left. \right|_{r\rightarrow r_S}=0 \; , \qquad e^{-\lambda} \left. \right|_{r\rightarrow r_S}=0 \; .
\end{equation}
The boundary condition for the scalar $\varphi$ at spatial infinity is given by Eq.~\eqref{scalarinfinity}.
To find the boundary condition for the scalar $\varphi$ at the horizon surface, one performs the expansion near the horizon: 
\begin{equation}
\label{expPhi}
e^{\nu}=a_1 (r-r_S)+...., \qquad e^{-\lambda}=b_1 (r-r_S)+..., \qquad \varphi=\varphi_S+\varphi'_S (r-r_S)+... \; ,
\end{equation}
where $a_1$ and $b_1$ are constants. 
Substituting Eq.~\eqref{expPhi} into Eq.~\eqref{tt} and keeping only the leading terms in $1/(r-r_S)$, we obtain
\begin{equation}
\label{b1}
b_1=  \frac{1-r^2_S V(\varphi_S)}{r_S-8\ell^2 \varphi_S \varphi'_S} \; .
\end{equation}
Now substituting expansions~\eqref{expPhi} into Eq.~\eqref{thetatheta}, we get
\begin{equation}
\label{thetathetahor}
\nu''+\frac{\nu'}2 (\nu'-\lambda') \left. \right |_{r \rightarrow r_S} \rightarrow  -\frac{1}{2r_S (r-r_S)} \cdot \left[1+8b_1 \ell^2 \varphi_S \varphi'_S +\frac{r_S}{b_1} V(\varphi_S)\right] \; .
\end{equation}
Finally, substituting Eqs.~(\ref{expPhi}) and (\ref{thetathetahor}) into Eq.~\eqref{eom}, we obtain 
\begin{equation}
\label{boundaryvarphiS}
\varphi'_S -\frac{16\ell^2 \varphi_S}{r^3_S} -\frac{8b_1 \ell^2\varphi_S}{r^2_S}  \left(1+\frac{16\ell^2 \varphi_S \varphi'_S}{r}   \right)  -\frac{1}{b_1} \left( V_{,\varphi} (\varphi_S) +\frac{16\ell^2 \varphi_S}{r^2_S}  V(\varphi_S)\right) =0 \; .
\end{equation}
Note that in the limit $V (\varphi) \rightarrow 0$, we consistently reproduce the result of Refs.~\cite{Antoniou:2017acq, Silva:2017uqg}.

We solve the system of Eqs.~\eqref{tt},~\eqref{rr}, and~\eqref{eom} with the boundary conditions~\eqref{scalarinfinity},~\eqref{boundarypotentials}, and~\eqref{boundaryvarphiS} numerically using both the shooting and relaxation methods. 
For the shooting technique we integrate from the horizon and use $\varphi_S$ as the shooting parameter. The value of $\varphi_S$ is adjusted such that the scalar field approaches the value $v$ at infinity. On the other hand, when using the relaxation method, one specifies the boundary condition~(\ref{boundaryvarphiS}) directly, while the correct value of $\varphi_S$ is found automatically within the method itself.

The results of numerical simulations are demonstrated in Figs.~\ref{scalarprofile},~\ref{metric}, and~\ref{metric2}, which show the scalar profile, and the metric functions $e^{\nu}$ and $e^{\lambda}$ relative to their GR values around Schwarzschild black holes, 
respectively. In particular, Fig.~\ref{scalarprofile} confirms the formation of walls in the large $\ell$ regime identified in the test field approximation, see Fig.~\ref{Test_field}. The presence of a wall has a sizeable impact on the metric leading to a considerable deviation from GR potentials even for $v \ll 1$. The deviation becomes more profound when the wall gets steep, while the parameters $\ell$ and $v$ are kept fixed -- this can be seen from the comparison of blue and green curves in Figs.~\ref{metric} and~\ref{metric2}. On the other hand, no walls 
are formed in the weak coupling regime $\ell \ll r_S$, yet large deviations from GR occur in this case  as well for substantially large values of $v$, i.e., $v \gtrsim 1$.

There is a limitation on the model parameters, for which regular static black hole solutions exist. For a large enough backreaction of the scalar field, the equations of motion become ill-defined at some $r$. 
More precisely, when the system of equations is written in a canonical form, the coefficient(s) of higher-order derivative variables are equal to zero at some $r$.
To see how the problem manifests, one eliminates $\nu''$ in Eq.~\eqref{eom} by making use of Eq.~\eqref{thetatheta}. As a result the equation for the field $\varphi$ takes the form
\begin{equation}
\left(1-\frac{128\ell^4 \varphi^2 \nu' e^{-\lambda} \cdot \left(1-e^{-\lambda} \right)}{r^2 \cdot \left[r+16\ell^2 e^{-\lambda} \varphi \varphi' \right]} \right) \varphi''+...= 0\; ,
\end{equation}
where by `...' we denote the terms with lower derivatives in $\varphi$.
Clearly, no regular solution exists, if the coefficient in front of the second derivative $\varphi''$ is zero at some $r$.
Using the Newtonian values of the gravitational potentials, i.e, assuming that the singularity in the equations of motion takes place reasonably far away from the horizon, 
one can express the condition for the existence of black hole solutions as follows: 
\begin{equation}
\frac{128 \ell^4 \varphi^2 r^2_S}{r^5 \cdot \left(r+16\ell^2 \varphi \varphi' \right)} \lesssim 1 \; .
\end{equation}
In the most interesting case of large $\ell$ and $\mu$, the quantity on the l.h.s. reaches the maximum at $r$ slightly larger than the crossover radius, where $\varphi \simeq v$, while the derivative 
$\varphi'$ is already negligible, i.e., $\varphi' \simeq 0$. Using these values and Eq.~\eqref{turn}, we arrive at the following constraint on the model parameters:  
\begin{equation}
v \lesssim \frac{0.4}{\mu \cdot \ell } \; .
\end{equation}
This constraint limits the backreaction of the field $\varphi$ on the gravitational potentials and hence departures from the GR metric, which nevertheless can be significant in the vicinity of the wall.

\section{Cosmology}
In this section, we briefly discuss the cosmological implications of the model given by Eqs.~\eqref{action},~\eqref{scalarpotential}, and~\eqref{coupling}. We show that consistency of the model with $\Lambda$CDM cosmology leads to important limitations on the early Universe evolution, 
in particular the reheating temperature $T_{\mathrm{reh}}$. Furthermore, possible problems at preheating as well as issues with domain walls and Dark Matter overproduction may invoke an extension of the model.  

{\it (In)stability in the early Universe.} On a cosmological background, the scalar-Gauss-Bonnet coupling feeds into the time-dependent mass of the field $\varphi$. This time-dependent mass can be tachyonic at certain epochs and reach very large values in the early Universe. As a result, one faces the risk of instability grossly 
altering the standard cosmological history~\cite{Anson:2019uto}. Indeed, in the FLRW Universe expanding with the scalar factor $a(t)$ the Gauss-Bonnet invariant can be written as 
\begin{equation}
\label{cosmoGB}
{\cal R}^2_{\mathrm{GB}}=24H^2 \frac{\ddot{a}}{a} \; ,
\end{equation}
where $H \equiv \dot{a}/a$ is the Hubble rate. Hence, for $\ell^2<0$ assumed in Refs.~\cite{Doneva:2017bvd, Silva:2017uqg} the field $\varphi$ has a positive effective mass squared in the decelerating Universe and tachyonic mass in the accelerating Universe. The latter fact has dramatic consequences for inflation, when expansion occurs with a nearly constant Hubble rate $H_{\mathrm{infl}}$ and the effective mass squared is given by 
\begin{equation}
\label{inflmass}
m^2_{\mathrm{infl}} =48 \ell^2 H^4_{\mathrm{infl}} \; .
\end{equation}
 Thus, for $\ell^2<0$ there is a huge instability, which is inconsistent with the existence of an inflationary stage\footnote{Note that the problem with instability during inflation is rather common for models predicting scalarization. In certain cases, e.g., in the model of Ref.~\cite{Damour:1993hw}, the problem can be cured by adding the coupling of the scalar responsible for scalarization to an inflaton~\cite{Anson:2019ebp}.}. Recall that formation of a scalar wall takes place for large values of $|\ell|$ greatly exceeding $H^{-1}_{\mathrm{infl}}$, unless one assumes an extremely low scale inflation.

In our case $\ell^2>0$, and the situation is reversed: the instability develops in the decelerating Universe, and vice versa. In particular, due to the large positive mass squared~\eqref{inflmass}, during inflation the field $\varphi$, including its perturbations, 
quickly relax to nearly zero values. This is true for any Fourier mode $\varphi (k, t)$ with wavenumbers $k/a \lesssim \ell\cdot H^2_{\mathrm{infl}}$, in which case the mode decays as $1/a^{3/2}$. So, for these modes by the end of inflation one estimates
\begin{equation}
\label{afterinflation}
\varphi (k, t_{\mathrm{e}}) \sim \frac{k}{\bar{a}}  \cdot \left( \frac{\bar{a}}{a_{\mathrm{e}}} \right)^{3/2} \sim \ell \cdot H^2_{\mathrm{infl}} \cdot \left(\frac{\bar{a}}{a_{\mathrm{e}}} \right)^{3/2} \; ,
\end{equation}
where $\bar{a}$ is the scale factor defined from $k/\bar{a} \sim \ell \cdot H^2_{\mathrm{infl}}$. For cosmologically interesting wavenumbers, these values $\varphi (k, t_{\mathrm{e}})$ are very small even given that $\ell \gg H^{-1}_{\mathrm{infl}}$. Note that we are not interested in modes with higher $k$, such that the inequality $k/a > l\cdot H^2_{\mathrm{infl}}$ is satisfied throughout inflation, since they are stable during the entire evolution of the Universe.

Despite this important advantage over the case with $\ell^2<0$, in our scenario the catastrophic instability may still take place during the radiation-dominated stage and possibly preheating, 
when the Universe is decelerating. Below we demonstrate this for the radiation-dominated stage and propose a solution to the problem. The equation of motion for the field $\varphi$ at the times of interest is given by 
\begin{equation}
\ddot{\varphi}+\frac{3 \dot{\varphi}}{2t} -\frac{3\ell^2}{t^4} \varphi =0 \; , 
\end{equation} 
where we have neglected the potential term and used $a(t)\propto t^{1/2}$.
The general solution of this equation is given by 
\begin{equation}
\varphi (t)=C_1 \cdot \left(\frac{\ell}{t} \right)^{1/4} \cdot  I_{1/4} \left(\frac{\sqrt{3} \ell}{t} \right) +C_2 \cdot \left(\frac{\ell}{t} \right)^{1/4} \cdot  K_{1/4} \left(\frac{\sqrt{3} \ell}{t} \right) \; ,
\end{equation}
where $I_{1/4}$ and $K_{1/4}$ are modified Bessel functions of the order $1/4$ of the first and second kind, respectively. We are interested in the growing part of the general solution $\sim C_2$. The coefficient $C_2$ is defined from the initial conditions for the field $\varphi$ and its derivative $\dot{\varphi}$ at the onset of the radiation-dominated stage, or the end of reheating, which we denote by $t_{\mathrm{reh}}$. At the times $t\gg t_{\mathrm{reh}}$ the solution for the field $\varphi$ reads 
\begin{equation}
\label{duringrd}
\varphi (t) \simeq \frac{1}{2}   \cdot \left(\frac{\ell}{t_{\mathrm{reh}}} \right)^{1/4} \cdot \left( \varphi_{\mathrm{reh}}  +\frac{\dot{\varphi}_{\mathrm{reh}} t^2_{\mathrm{reh}}}{\sqrt{3}l} \right) \cdot e^{\frac{\sqrt{3}\ell}{t_{\mathrm{reh}}}} \cdot K_{1/4} \left(\frac{\sqrt{3} \ell}{t} \right)  \; ,
\end{equation}
where we assumed $\ell/t_{\mathrm{reh}}\gg 1$. As it follows, even for tiny initial $\varphi_{\mathrm{reh}}$ and $\dot{\varphi}_{\mathrm{reh}}$ and generically large $\ell/t_{\mathrm{reh}}$, the field $\varphi$ almost instantly reaches huge values, which are fatal for the existence of the radiation-dominated stage. 

One obvious solution to this problem is to ensure that the exponent in Eq.~(\ref{duringrd}) is cancelled by the exponentially suppressed amplitude of the scalar at $t_{\mathrm{reh}}$, i.e., to assume
that $\ell/t_{\mathrm{reh}} \lesssim \ln \left(M_{\mathrm{Pl}}/\varphi_{\mathrm{reh}} \right) \lesssim 100$. (In this section, we do not set $8\pi G_N=1$, and restore the Planck mass $M_{\mathrm{Pl}} \approx 2.44 \cdot 10^{18}~\mbox{GeV}$.) The latter inequality here is fulfilled provided that the field $\varphi$, strongly decaying during inflation, 
does not experience a significant growth at preheating (see the comment at the end of this paragraph)\footnote{With this assumption, one has $\varphi_{\mathrm{reh}} (k) \sim \varphi (k, t_{\mathrm{e}})$. To estimate $\varphi (k, t_{\mathrm{e}})$ from Eq.~\eqref{afterinflation} and hence $\varphi_{\mathrm{reh}}$, we note that only the modes with the wavenumber $k/a_{\mathrm{reh}} \sim \ell\cdot H^2_{\mathrm{reh}}$ or lower may experience instability during the radiation-dominated stage. Such modes exponentially decay during inflation starting from the 
moment, when $\frac{k}{\bar{a}} \sim \ell \cdot H^2_{\mathrm{infl}}$. Hence, for the unstable mode with the maximum possible wavenumber one has $\bar{a} \sim (H_{\mathrm{reh}}/H_{\mathrm{infl}})^2 \cdot a_{\mathrm{reh}} $. The modes with lower $k$, which are also unstable, have smaller initial amplitudes; therefore we concentrate on the particular mode satisfying $k/a_{\mathrm{reh}} \sim \ell\cdot H^2_{\mathrm{reh}}$. Consequently, using Eq.~\eqref{afterinflation}, we find:
\begin{equation}
\varphi_{\mathrm{reh}} \sim \ell H_{\mathrm{reh}}^2 \cdot \left(\frac{H_{\mathrm{reh}}}{H_{\mathrm{infl}}} \right) \cdot \left(\frac{a_{\mathrm{reh}}}{a_{\mathrm{e}}} \right)^{3/2} \; .
\end{equation}
Taking the values $H_{\mathrm{infl}} \sim 10^{13}~\mbox{GeV}$, $T_{\mathrm{reh}} \simeq 1~\mbox{GeV}$, assuming $\ln (\ell H_{\mathrm{reh}}) <100 $ and not extremely large $a_{\mathrm{reh}}/a_{\mathrm{e}}$, we get $\ln (M_{\mathrm{Pl}}/\varphi_{\mathrm{reh}}) \sim 100$ -- the value quoted in the main text.}.
The condition $\ell/t_{\mathrm{reh}} \lesssim 100$ has strong consequences for the reheating temperature of the Universe $T_{\mathrm{reh}}$. To demonstrate this, we use the expression for the Hubble rate during the radiation-dominated stage:
\begin{equation}
\label{hubblerd}
H (T)=\sqrt{\frac{\pi^2 g_* (T)}{90}} \cdot \frac{T^2}{M_{\mathrm{Pl}}} \; ,
\end{equation}
where $g_* (T)$ is the number of relativistic degrees of freedom. On the other hand, the Hubble rate at radiation-domination is given by $H=1/(2t)$. Equating the latter to the expression~\eqref{hubblerd} and using $\ell/t_{\mathrm{reh}} \lesssim 100$, we obtain from Eq.~\eqref{hubblerd}: 
\begin{equation}
T_{\mathrm{reh}} \lesssim 10 \cdot \left(\frac{90}{\pi^2 g_* (T_{\mathrm{reh}})} \right)^{1/4} \cdot \left(\frac{M_{\mathrm{Pl}}}{\ell} \right)^{1/2} \; .
\end{equation}
For the interesting values of $\ell$ of the order of the Schwarzschild radius of the Sun, the above constraint reads
\begin{equation}
\label{reh}
T_{\mathrm{reh}} \lesssim 3~\mbox{GeV} \; .
\end{equation}
We have set $g_* (T_{\mathrm{reh}}) \simeq 30$. Note that the upper bound on $T_{\mathrm{reh}}$ here is completely reasonable: it is three orders of magnitude above the lower bound, $T_{\mathrm{reh}} > 4~\mbox{MeV}$ at $95\%$ CL~\cite{Hannestad:2004px}. 
Still, such low reheating temperatures imply that there is a long preheating stage separating the inflationary and radiation dominated eras. As the equation of state during 
preheating corresponds to deceleration, one again expects a runaway solution for the field $\varphi$. Whether or not this runaway can be achieved in a controllable way, remains 
an open issue.  

Let us make one comment in passing. The constraint~\eqref{reh} is valid independently of a particular choice of the potential $V(\varphi)$ and only assumes $\ell^2>0$. 
Namely, it is also applicable in the model with $V(\varphi)=0$. While the latter scenario requires $\ell^2<0$ for scalarization of static black holes, it has been shown in Refs.~\cite{Dima:2020yac, Berti:2020kgk, Herdeiro:2020wei} that rapidly rotating Kerr black holes can develop scalar hair also in the case $\ell^2>0$.

{\it Domain walls.} Formation of domain walls is common in models exhibiting spontaneous breaking of discrete symmetries~\cite{Zeldovich:1974uw}, which is the case we are considering. As discussed above, the field $\varphi$ quickly relaxes to zero during inflation. Thus, during the subsequent evolution the field $\varphi$ occupies vacua with positive and negative expectation values, i.e., $\varphi =\pm v$ in causally disconnected patches with equal probability. Neighbouring regions, where the field $\varphi$ takes opposite expectation values, are separated by domain walls. Typically, the latter pose problems for $\Lambda$CDM model, because 
they tend to overclose the Universe and affect cosmic microwave background measurements in unacceptable manner. These considerations set a strong limit on the domain wall tension~\cite{Zeldovich:1974uw, Lazanu:2015fua}:
\begin{equation}
\sigma =\frac{2\sqrt{2} h v^3}{3} \lesssim (1~\mbox{MeV})^3 \; ,
\end{equation}
which can be converted into 
\begin{equation}
\label{toorestrictive}
\mu \lesssim 0.2 \cdot 10^{-36} \cdot \left(\frac{M_{\mathrm{Pl}}}{v} \right)^2~\mbox{eV} \; .
\end{equation}
For example, assuming the expectation values $v \simeq 0.1~M_{\mathrm{Pl}}$, which are relevant from the viewpoint of shell formation, we obtain $\mu \lesssim 2 \cdot 10^{-33}~\mbox{eV}$. 
With such a strong limit, the potential $V(\varphi)$ can be neglected for the study of black holes of astrophysical size. This is not an option of our major interest in the present work. The restrictive bound~\eqref{toorestrictive} can be avoided, if one allows for a small explicit breaking of $Z_2$-symmetry, in which case domain walls are dissolved at some point of cosmological history~\cite{Zeldovich:1974uw, Gelmini:1988sf}. 
We assume that the small breaking of $Z_2$-symmetry does not affect the analysis of black hole solutions, but this remains to be shown explicitly.

{\it Dark Matter.} Note that the field $\varphi$ generically experiences oscillations around a minimum of the broken phase. The oscillations contribute to Dark Matter, and this suggests another way of constraining the model. 
They start at the time $t_*$, when $\mu \simeq H_*$. Using Eq.~\eqref{hubblerd}, we obtain for the temperature of the Universe $T_*$ at this moment:
\begin{equation}
\label{temposc}
T_* \simeq \left(\frac{90}{\pi^2 g_* (T_{*})} \right)^{1/4} \cdot \sqrt{\mu M_{\mathrm{Pl}}} \; .
\end{equation}
At the onset of oscillations, their amplitude is naturally estimated as $\mu/h$ and then redshifts $\propto 1/a^{3/2}$. Thus, we can estimate the energy density of the Dark Matter component due to the field $\varphi$ as 
\begin{equation}
\rho_{\mathrm{osc}} (t)\simeq \frac{\mu^4}{h^2} \cdot \left(\frac{a_*}{a(t)} \right)^3 \; .
\end{equation}
The requirement that Dark Matter is not overproduced can be expressed as $\rho_{\mathrm{osc}} (t_{\mathrm{eq}}) \lesssim \rho_{\mathrm{rad}} (t_{\mathrm{eq}})$, where $t_{\mathrm{eq}}$ is the time of the matter-radiation equality, and 
$\rho_{\mathrm{rad}} (t_{\mathrm{eq}})$ is the radiation energy density at this time. Using this inequality, entropy conservation in the comoving volume, which gives
\begin{equation}
\left(\frac{a_*}{a_{\mathrm{eq}}} \right)^3 \simeq \frac{g_* (T_{\mathrm{eq}}) \cdot T^3_{\mathrm{eq}}}{g_* (T_*) \cdot T^3_*} \; , 
\end{equation}
and Eq.~\eqref{temposc}, expressing the quartic self-interaction constant $h^2$ through the expectation value $v=\mu/h$, we obtain 
\begin{equation}
\mu \lesssim 2  \cdot \left(\frac{M_{\mathrm{Pl}}}{v} \right)^4  \cdot 10^{-27}~\mbox{eV} \; ,
\end{equation}
where we have set $g_* (T_*) \approx 3.4$. 
The above bound would place a strong constraint on the range of parameters of our model. One can avoid the problem by adding a decay channel for oscillations of the field $\varphi$ into lighter degrees of freedom. We postpone further discussion of the problem for future studies.

{\it Speed of gravitational waves.} Finally, let us ensure that our model is consistent with the observations of the gravitational wave signal GW170817~\cite{LIGOScientific:2017vwq} and the complementary gamma ray burst~\cite{Goldstein:2017mmi}, which constrain 
the speed of gravitational waves to be $|c_T-1| \lesssim 10^{-15}$~\cite{LIGOScientific:2017zic} disfavouring a number of dark energy and modified gravity scenarios~\cite{Baker:2017hug, Ezquiaga:2017ekz, Creminelli:2017sry}. In our model, the speed of gravitational waves is given by~\cite{Tattersall:2018map} 
\begin{equation}
c_T=1 +16\ell^2 \frac{H \dot{\varphi} \varphi-\ddot{\varphi}  \varphi-\dot{\varphi}^2}{M^2_{\mathrm{Pl}}-16\ell^2 H \dot{\varphi} \varphi} \; .
\end{equation}
The field $\varphi$ is assumed to be in the minimum of spontaneously broken phase slowly drifting due to the Gauss-Bonnet coupling: 
\begin{equation}
\varphi \approx v \cdot \left(1-\frac{\ell^2}{\mu^2} {\cal R}^2_{\mathrm{GB}} \right) \; . 
\end{equation}
Using Eq.~\eqref{cosmoGB}, we estimate 
\begin{equation}
|\dot{\varphi}| \sim \frac{100 \ell^2 H^5 v }{\mu^2}  \; ,
\end{equation}
and $|\ddot{\varphi}| \sim H |\dot{\varphi}|$. The speed of gravitational waves is measured in the late Universe, when the Hubble rate is extremely small in the range of parameters interesting for black hole physics, i.e., 
$H \ll \ell^{-1}$ and $H \ll \mu$. Hence, we have $\dot{\varphi}^2 \ll H \varphi |\dot{\varphi}| \sim \varphi |\ddot{\varphi}|$ and thus 
\begin{equation}
|c_T-1| \sim \frac{1000 \ell^4 H^6}{\mu^2} \cdot \frac{v^2}{M^2_{\mathrm{Pl}}} \; .
\end{equation}
Taking, e.g., $\ell \sim \mu^{-1} \sim r_S \sim 3~\mbox{km}$ (Schwarzschild radius of the Solar mass object), $v \sim 0.1 M_{\mathrm{Pl}}$, and the present day Hubble rate $H_0 \approx 70~\mbox{km}/(s \cdot \mbox{Mpc})$, we obtain 
\begin{equation}
\label{cGW}
|c_T -1| \sim 10^{-135}.
\end{equation}
Note that the speed of gravity in the vicinity of a  black hole deviates from the speed of light by a greater amount, however, the region of space where this happens is negligibly  small in comparison to the distance from the source of the event GW170817.
We conclude that for any reasonable values of $\ell,~\mu$, and $v$, the speed of gravitational waves is well within the existing observational bound\footnote{Note that claiming the exact equality $c_T=1$ in a model with a scalar-Gauss-Bonnet coupling leads 
to a conflict with the early Universe cosmology, e.g., with inflation~\cite{Linder:2021pek}. We have shown that the conflict is avoided, once the constraint $c_T=1$ is relaxed by just a tiny bit.}.

\section{Discussions}

In the present work we have constructed static black hole solutions in a model described by Eqs.~\eqref{action},~\eqref{scalarpotential}, and~\eqref{coupling}. 
Introducing the potential $V(\varphi)$ is natural in the quadratic scalar-Gauss-Bonnet model, as the latter does not respect shift symmetry. Compared to previous studies, we considered the case in which the potential $V(\varphi)$ spontaneously breaks $Z_2$-symmetry, so that the scalar $\varphi$ acquires a non-zero 
vacuum expectation value. Such a choice of potential is very common in particle physics, with the Higgs field being the most well-known example. 

For $\ell \ll r_S$, the solution is in the perturbative regime, meaning that the departure of the scalar field $\varphi$ from its expectation value is small and goes to zero in the limit $\ell^2 \rightarrow 0$. While the effect of the Gauss-Bonnet coupling is rather mild in this regime, the case of small $\ell$ is nevertheless interesting at least for two reasons. First, it allows for the perturbative treatment of black hole mergers, cf.~Ref.~\cite{Witek:2018dmd}. Second, for $\ell \ll r_S$ we effectively recover the linear Gauss-Bonnet coupling $\sim v \varphi  {\cal R}^2_{\mathrm{GB}}$. Remarkably, the recovery occurs
in the regime, where the linear coupling can be trusted according to the recent causality constraints~\cite{Serra:2022pzl}. Note that the coupling $\sim v \varphi {\cal R}^2_{\mathrm{GB}}$ is commonly discussed in the context of shift-symmetric models, while in our case the shift-symmetry is explicitly broken by the potential $V(\varphi)$.

For $\ell \gg r_S$, which is the most interesting case, the scalar profile around the black hole depends non-perturbatively on the coupling constant $\ell^2$. The $Z_2$-symmetry, spontaneously broken at spatial infinity, is restored in the vicinity of the black hole,
where the scalar field takes on exponentially small values.
This gives rise to the main effect identified in this work: a black hole is shrouded by a wall for sufficiently 
large masses $\mu$. Namely, the scalar field experiences a steep growth in a thin shell separating the domains, where it is in the broken phase and in the unbroken phase. This effect is already manifested in the test field approximation (see Figs.~\ref{Test_field} and~\ref{Derivative_test}) and holds upon taking dynamics of the metric into account (see Fig.~\ref{scalarprofile}). 
As it can be seen from Figs.~\ref{metric} and~\ref{metric2}, the thinner the wall is, the larger the deviation from the Schwarzschild metric is for a fixed coupling constant $\ell^2$ and expectation value $v$. 
Checking stability of black hole solutions we obtained in this paper, deserves a separate investigation in the line of Refs.~\cite{Minamitsuji:2018xde, Silva:2018qhn}. In the future, it would be also interesting to see how the walls are manifested in the case of Kerr black holes and neutron stars. Furthermore, it is worth investigating implications of the walls for strong field tests of GR: black hole mergers and binary pulsars.  

Finally, let us mention that besides black hole physics, early Universe cosmology represents a promising playground for studying GR in the large curvature regime. In the present work we briefly considered cosmological implications 
of the model at hand in Section~5. Unlike the commonly considered scenario of spontaneous scalarization with $\ell^2<0$, no problem with catastrophic tachyonic instability at inflation arises in our case. The instability appears during the radiation-dominated stage, but it is harmless provided that the reheating temperature is constrained to be sufficiently low, $T_{\mathrm{reh}} \lesssim 1~\mbox{GeV}$. As a result, one expects a rather long stage of preheating, which can be highly non-trivial due to the presence of the scalar-Gauss-Bonnet coupling and thus worth investigating. In addition, consistency with the $\Lambda$CDM model must be carefully studied in view of possible issues with Dark Matter overproduction and formation of domain walls. These challenges may necessitate a modification of our scenario. We leave this task for future work. It is also worthwhile noting, that in our model the speed of gravitational waves is extremely close to the speed of light, Eq.~(\ref{cGW}), being well within the present bound from the event GW170817 and its complimentary gamma ray burst.

\section*{Acknowledgments}

S.R. is indebted to Leonardo Trombetta for useful discussions. The work of W.~E. and S.~R. is supported by the Czech Science Foundation GA\v CR, project 20-16531Y.

\section*{Appendix. Analytical solutions in test field approximation} 

In this Appendix, we construct an approximate analytical solution of Eq.~\eqref{testequation} with the boundary conditions~\eqref{scalarinfinity} and~\eqref{testboundary}. 
First, we notice that in the range of masses $\mu$ interesting for our scenario and for $\varphi_S \lesssim v$, we have
\begin{equation}
\label{simplification}
|V_{,\varphi} | \lesssim \mu^2 \varphi_S \ll \frac{25 \ell^2}{r^4_S} \varphi_S\; .
\end{equation}
Thanks to the above inequality the last term in Eq.~\eqref{testboundary} can be dropped, so that the latter simplifies to  
\begin{equation}
\label{testboundarys}
 \varphi'_S -\frac{24\ell^2}{r^3_S} \cdot \varphi_S =0\; ,
\end{equation}
which we assume in what follows.

For generic model parameters, there are three separate ranges of $r$, where the solution for the field $\varphi$ has a distinct behaviour: i) close to the horizon, ii) intermediate range, i.e., far from the horizon but inside the crossover radius given by Eq.~(\ref{turn}), iii) above the crossover radius.
For each range of $r$ we find the general analytical solution, and then construct a full solution by matching the analytical solutions on the boundaries between the regions.

In the region of small $r$, i.e., close to the black hole horizon,
\begin{equation}
r-r_S \ll r_S \; ,
\end{equation}
Eq.~\eqref{testequation} can be written as
\begin{equation}
\varphi''+\frac{\varphi'}{r-r_S} -\frac{24 \varphi \ell^2}{r^3_S \cdot (r-r_S)} =0 \; ,
\end{equation}
where we again took into account the inequality~\eqref{simplification}. The solution of the above equation respecting the boundary condition~\eqref{testboundarys} reads
\begin{equation}
\label{veryclose}
\varphi = \varphi_S \cdot I_0 \left(\frac{\sqrt{96} \ell}{r_S} \cdot \sqrt{\frac{r}{r_S}-1} \right) \; ,
\end{equation}
where $I_0$ is the modified Bessel function of the first kind of the zeroth order. 

In the intermediate region
\begin{equation}
r_S \ll r \ll r_{\mathrm{cross}} \; ,
\end{equation}
one can replace the gravitational potential $\nu$ in Eq.~\eqref{testequation} by its Newtonian value $\nu\simeq -r_S/r$ and neglect the potential term with respect to the Gauss-Bonnet term using~(\ref{turn}). As a result, we obtain
\begin{equation}
\varphi''+\frac{2}{r} \varphi' -\frac{24\ell^2 r^2_S}{r^6} \varphi =0 \; .
\end{equation}
The general solution of this equation reads 
\begin{equation}
\label{testinter}
\varphi =\frac{1}{2^{1/4}}\left(\frac{\sqrt{6} \ell r_S}{r^2} \right)^{1/4} \left[ \frac{2 C_1}{\Gamma \left(\frac{1}{4} \right)}  K_{1/4} \left(\frac{\sqrt{6} \ell r_S}{r^2} \right)+C_2  \Gamma \left(\frac{5}{4} \right) I_{1/4} \left(\frac{\sqrt{6} \ell r_S}{r^2} \right)\right] \; ,
\end{equation}
where $\Gamma$ is a Gamma function; $I_{1/4}$ and $K_{1/4}$ are modified Bessel functions of the order $1/4$ of the first and second kinds, respectively. 

Finally, at very large distances
\begin{equation}
r \gg r_{\mathrm{cross}} \; , 
\end{equation}
the equation of motion for the field $\varphi$ is given by
\begin{equation}
\tilde{\varphi}''+\frac{2}{r} \tilde{\varphi}'-2\mu^2 \tilde{\varphi}=0 \; ,
\end{equation}
where $\tilde{\varphi} \equiv \varphi-v$ and the Gauss-Bonnet term was neglected as compared to the potential term. 
With the boundary condition $\varphi \rightarrow v$ in the limit $r \rightarrow \infty$, one gets from the above equation:
\begin{equation}
\label{testouter}
\varphi=v+\frac{D e^{-\sqrt{2} \mu r}}{r} \; .
\end{equation}
The general expressions~\eqref{veryclose},~\eqref{testinter}, and~\eqref{testouter} contain arbitrary constants $C_1,~C_2$, and $D$ as well as the value of the scalar field at the horizon $\varphi_S$. 
They are fixed by matching the solutions for different ranges of $r$. In what follows, we treat separately three cases outlined in Eq.~\eqref{cases}.

{\it Case~I}. In this case, the arguments of the modified Bessel functions are small in the intermediate region $r_{S} \ll r \ll r_{\mathrm{cross}}$. Therefore, we can use the following limiting expressions:
\begin{equation}
\label{lowargument}
I_{1/4} (z \ll 1) \approx \frac{1}{\Gamma \left(\frac{5}{4} \right)} \cdot \left(\frac{z}{2} \right)^{1/4} \qquad K_{1/4} (z \ll 1) \approx \frac{1}{2} \Gamma \left(\frac{1}{4} \right) \cdot \left(\frac{2}{z} \right)^{1/4} +\frac{1}{2} \Gamma \left(-\frac{1}{4} \right) \cdot
\left(\frac{z}{2} \right)^{1/4} \; ,
\end{equation}
and rewrite the general solution~\eqref{testinter} as
\begin{equation}
\label{smalll}
\varphi \approx C_1 +\frac{6^{1/4}}{\sqrt{2}} \frac{\sqrt{\ell r_S}}{r} \cdot \left[\frac{\Gamma \left(-\frac{1}{4} \right)}{\Gamma \left(\frac{1}{4} \right)} \cdot C_1 +C_2 \right] \; . 
\end{equation}
Matching the latter to Eq.~\eqref{testouter} at $r \sim r_{\mathrm{cross}}$ gives 
\begin{equation}
\label{C1}
C_1 \simeq v \; ,
\end{equation}
and 
\begin{equation}
\label{D}
D \simeq 6^{1/4} \cdot \sqrt{\frac{ \ell r_S}{2}} \cdot \left[\frac{\Gamma \left(-\frac{1}{4} \right)}{\Gamma \left(\frac{1}{4} \right)} \cdot v+C_2 \right] \; .
\end{equation}
Now using the near horizon solution~\eqref{veryclose} and the small argument expansion of the modified Bessel function $I_0$: 
\begin{equation}
\label{modBzerosmall}
I_0 (z \ll 1) \approx 1+\frac{z^2}{4} \; ,
 \end{equation}
 we estimate the coefficient $C_2$:
\begin{equation}
\label{C2smalll}
\frac{\Gamma \left(-\frac{1}{4} \right)}{\Gamma \left(\frac{1}{4} \right)} v+C_2 \simeq -\frac{24 \cdot \sqrt{2} \cdot \ell^{3/2} \cdot r^2_* \cdot v}{6^{1/4} \sqrt{r_S} \cdot \left[r^3_S+24\ell^2 \cdot (2r_* -r_S) \right]}  \sim 10 v \left( \frac{\ell}{r_S} \right)^{3/2} \; ,
\end{equation}
where $r_*$ is the distance where the near-horizon asymptotic behaviour of the scalar~(\ref{veryclose}) transitions to the intermediate type, Eq.~(\ref{testinter}).
In other words, it is the distance at which the Newtonian approximation in Eq.~\eqref{testequation} becomes valid.
Typically $r_*>r_S$ is of the order of a Schwarzschild radius, $r_* ={\cal O} (1) r_S$.
We obtain the field value $\varphi_S$ at the horizon surface: 
\begin{equation}
\label{initialI}
\varphi_S \simeq \frac{v\cdot r^3_S}{r^3_S+24 \ell^2  \cdot \left(2r_*-r_S \right)}   \sim v \left[1- 10  \left( \frac{\ell}{r_S} \right)^{2} \right] \; .
\end{equation}
Substituting Eq.~\eqref{C2smalll} into Eq.~\eqref{D}, we obtain
\begin{equation}
\label{Dsmall}
D \simeq -\frac{24 \ell^2 r^2_* v}{r^3_S +24\ell^2 \cdot (2r_*-r_S) }  \sim -10 \frac{\ell^2 v}{r_S} \; .
\end{equation}
Substituting Eqs.~\eqref{C2smalll} and~\eqref{Dsmall} into Eqs.~\eqref{testinter} and~\eqref{testouter}, respectively, we can write the final solution in the unified form~\eqref{profilecaseI}.

{\it Case~II}. In this case, the expressions~\eqref{C1} and~\eqref{D} hold, since the expansion~(\ref{lowargument}) is also valid in this case at $r\sim r_{\mathrm{cross}}$. 
The difference with the previous case is that close to the horizon Eqs.~(\ref{lowargument}) and~\eqref{modBzerosmall} are not valid any longer, since the arguments of the modified Bessel functions become large.  One can use instead the following asymptotic expressions:
\begin{equation}
\label{largeargument}
I_{1/4} (z \gg 1) \sim \frac{e^{z}}{\sqrt{2\pi z}} \; , \qquad K_{1/4} (z \gg 1) \sim \sqrt{\frac{\pi}{2z}} e^{-z}\; ,
\end{equation}
and
\begin{equation}
\label{largeargmodB}
I_0 (z \gg 1) \approx \frac{e^{z}}{\sqrt{2\pi z}} \; .
\end{equation}
Then, we match the solution~\eqref{testinter}, where we use Eq.~\eqref{largeargument}, with the near-horizon solution~\eqref{veryclose}, where we use Eq.~(\ref{largeargmodB}), at the radius $r_*$. As in case I, the radius $r_*$ is of the order of a Schwarzschild radius. We find that both $\varphi_S$ and $C_2$ are exponentially suppressed, i.e.,
\begin{equation}
\label{C2largel}
C_2  \sim e^{-\frac{2\sqrt{6} \ell r_S}{r^2_*}}\; ,
\end{equation}
and 
\begin{equation}
\label{initialII}
\varphi_S \sim e^{-\frac{\sqrt{6}\ell r_S}{r^2_*} -\frac{4\sqrt{6} \ell}{r_S} \cdot \sqrt{\frac{r_*}{r_S}-1}}\; .
\end{equation}
Given large exponential suppression, one can neglect the term $\sim C_2$ in Eq.~\eqref{testinter}. As a result, we get in the intermediate regime $r_S \ll r \ll r_{\mathrm{cross}}$:
\begin{equation}
\label{case2inter}
\varphi \simeq \frac{2^{3/4} \cdot v}{\Gamma \left(\frac{1}{4} \right)} \cdot \left(\frac{\sqrt{6} \ell r_S}{r^2} \right)^{1/4} \cdot K_{1/4} \left(\frac{\sqrt{6} \ell r_S}{r^2} \right) \; .
\end{equation}
For not very large $r$, i.e., $r_S \ll  r \ll \sqrt{\ell r_S}$, the latter expression reduces to Eq.~\eqref{profileinterII} from the main text, which describes exponential suppression. For $\sqrt{\ell r_S} \ll  r \ll r_{\mathrm{cross}}$, Eq.~\eqref{case2inter} takes the form~(\ref{smalll}) with $C_1 \simeq v$ and  $C_2 \simeq 0$, while for larger $r$, i.e., $r\gg r_{\mathrm{cross}}$, Eq.~(\ref{testouter}) is valid with 
\begin{equation}
D \simeq \left(\frac{3}{2} \right)^{1/4}  \cdot \frac{\Gamma \left(-\frac{1}{4} \right)}{\Gamma \left(\frac{1}{4} \right)}\cdot \sqrt{\ell r_S} \cdot v\; ,
\end{equation}
which can be inferred from Eq.~(\ref{D}) by setting $C_2=0$. The two expressions for $r\gg \sqrt{\ell r_S}$ can be written in the unified form~\eqref{profilelargeII}.

{\it Case III.} In this case, the near horizon behaviour of the field $\varphi$ is similar to the case above. 
In particular, the coefficient $C_2$ is exponentially suppressed. More precisely, 
it is given by Eq.~\eqref{C2largel} with the replacement $v \to C_1$. 
Thus, again we can set to zero $C_2$ in Eq.~\eqref{testinter}  in the intermediate regime. 
Unlike case II, however, in case III, the asymptotic expressions~\eqref{largeargument} for the modified Bessel functions are valid in the whole intermediate range 
including the crossover radius $r_{\mathrm{cross}}$. The coefficients $C_1$ and $D$ are obtained by matching expressions~\eqref{testinter} and~\eqref{testouter} at $r_{\mathrm{cross}}$ in a straightforward manner. We get
\begin{equation}
\label{C1large}
C_1 \simeq \frac{\Gamma \left(\frac{1}{4} \right)}{2^{1/4} \sqrt{\pi}} \cdot \left(\frac{\sqrt{6} \ell r_S}{r^2_{\mathrm{cross}}} \right)^{1/4} \cdot \frac{\mu r^3_{\mathrm{cross}}}{\mu r^3_{\mathrm{cross}} +2\sqrt{3} \ell r_S} \cdot e^{\frac{\sqrt{6} \ell r_S}{r^2_{\mathrm{cross}}}} \cdot v\; ,
\end{equation}
and
\begin{equation}
\label{Dlarge}
D \simeq -\frac{2\sqrt{3}\ell r_S r_{\mathrm{cross}} }{\mu r^3_{\mathrm{cross}}+2\sqrt{3} \ell r_S} \cdot e^{\sqrt{2} \mu r_{\mathrm{cross}}}\cdot v \; .
\end{equation}
Substituting Eq.~\eqref{C1large} into Eq.~\eqref{testinter}, where we omit the term $\sim C_2$, and using Eq.~\eqref{turn} we obtain Eq.~\eqref{profileinterIII}. 
Substituting Eq.~\eqref{Dlarge} into Eq.~\eqref{testouter}, and using Eq.~\eqref{turn} we obtain Eq.~\eqref{profilelargeIII}. 
The horizon value $\varphi_S$ is obtained by matching Eq.~\eqref{veryclose} to Eq.~\eqref{testinter}, where $C_1$ is given by Eq.~\eqref{C1large} and $C_2$ is obtained from Eq.~\eqref{C2largel} upon replacing $v$ by $C_1$; we also use Eq.~\eqref{largeargmodB}. Again the resulting value $\varphi_S$ is exponentially suppressed: 
\begin{equation}
\varphi_S \sim e^{\sqrt{6} \ell r_S \cdot \left(\frac{1}{r^2_{\mathrm{cross}}} -\frac{1}{r^2_*}\right) -\frac{4\sqrt{6} \ell}{r_S} \cdot \sqrt{\frac{r_*}{r_S}-1}} \; ,
\end{equation}
but the suppression is milder than in case II, cf. Eq.~\eqref{initialII}.
This reflects the simple fact that the coupling to the Gauss-Bonnet invariant becomes less efficient as one increases $\mu$, and the field $\varphi$ tends to stay closer to its expectation value $v$.

Finally, let us discuss the behaviour of the field $\varphi$ at the wall location close to $r_{\mathrm{cross}}$ in case III. The relevant equation of motion is obtained by simplifying Eq.~\eqref{testequation}, which reads in the Newtonian approximation: 
\begin{equation}
\label{nearwall}
\varphi''+\frac{2}{r} \varphi' -\frac{24 l^2 r_S}{r^6} \varphi -h^2 \varphi (\varphi^2 -v^2)=0 \; .
\end{equation}
Assuming that the wall width is small relative to the crossover radius, i.e., $\delta_{\mathrm{wall}} \ll r_{\mathrm{cross}}$ one can neglect the second term on the r.h.s., since $|\varphi'/r | \ll |\varphi''|$ in that case. 
Somewhat loosely we also drop the cubic term $\sim h^2 \varphi^3$, and Eq.~\eqref{nearwall} takes the form 
\begin{equation}
\label{wallcentersimple}
\varphi''+\left(\mu^2 -\frac{24l^2 r^2_S}{r^6} \right) \varphi =0 \; . 
\end{equation}
Note that the coefficient in front of the second term here turns into zero at $r_{\mathrm{cross}}$, so that one can perform the expansion:
\begin{equation}
\mu^2 -\frac{24l^2 r^2_S}{r^6} \simeq \frac{144l^2 r^2_S}{r^7_{\mathrm{cross}}} \cdot (r-r_{\mathrm{cross}})\; .
\end{equation}
Consequently, Eq.~\eqref{wallcentersimple} can be written as 
\begin{equation}
\label{nearwallverysimple}
\frac{d^2 \varphi}{dx^2} + x \varphi =0\; ,
\end{equation}
where 
\begin{equation}
\label{x}
x \equiv \left(\frac{144l^2 r^2_S}{r^7_{\mathrm{cross}}} \right)^{1/3} \cdot (r-r_{\mathrm{cross}}) \; .
\end{equation}
The general solution of Eq.~\eqref{nearwallverysimple}, which can be written as a linear combination of Airy functions, is not particularly illuminating. Without invoking the solution, however, one can deduce that
there is only one dimensionful parameter in this regime given by the coefficient in front of $(r-r_{\mathrm{cross}})$ in Eq.~\eqref{x}. This coefficient is thus identified 
as the inverse wall width, i.e.,   
\begin{equation}
\delta_{\mathrm{wall}} \sim \left(\frac{r^7_{\mathrm{cross}}}{144l^2 r^2_S} \right)^{1/3} \; .
\end{equation}
Using Eq.~\eqref{turn}, we get the estimate~\eqref{width} from the main text.

\end{document}